\documentclass[%
twocolumn,
showpacs,
nofootinbib,
 amsmath,amssymb,
 aps,
]{revtex4}
\usepackage{graphicx}
\usepackage{dcolumn}
\usepackage{bm}
\usepackage{color}
\usepackage{float}
\usepackage{braket}
\usepackage{amsmath}
\usepackage{verbatim}
\usepackage{multirow}

\newcommand{\pd}{\partial}

\def\tagform@#1{\maketag@@@{\bfseries(\ignorespaces#1\unskip\@@italiccorr)}}
\renewcommand{\eqref}[1]{\textup{{\normalfont(\ref{#1}}\normalfont)}}

\begin{document}

\title{High harmonic generation from Bloch electrons in solids}

\author{Mengxi Wu$^1$}
\email{mwu3@lsu.edu}
\author{Shambhu Ghimire$^2$}
\author{David A. Reis$^{2,3}$}
\author{Kenneth J. Schafer$^1$}
\author{Mette B. Gaarde$^1$}
\email{gaarde@phys.lsu.edu}

\affiliation{1) Department of Physics and Astronomy, Louisiana State University, Baton Rouge, LA 70803-4001, USA}
\affiliation{2) PULSE Institute, SLAC National Accelerator Laboratory, Menlo Park, California, 94025, USA}
\affiliation{3) Departments of Photon Science and Applied Physics, Stanford University, Stanford, California, 94305, USA}
\date{\today}

\begin{abstract}

We study the generation of high harmonic radiation by Bloch electrons in a model transparent solid driven by a strong mid-infrared laser field. We solve the single-electron time-dependent Schr\"odinger equation (TDSE) using a velocity-gauge method [New J. Phys. 15, 013006 (2013)] that is numerically stable as the laser intensity and number of energy bands are increased. The resulting harmonic spectrum exhibits a primary plateau due to the coupling of the valence band to the first conduction band, with a cutoff energy that scales linearly with field strength and laser wavelength. We also find a weaker second plateau due to coupling to higher-lying conduction bands, with a cutoff that is also approximately linear in the field strength.  To facilitate the analysis of the time-frequency characteristics of the emitted harmonics, we also solve the TDSE in a time-dependent basis set, the Houston states [Phys. Rev. B 33, 5494 (1986)], which allows us to separate inter-band and intra-band contributions to the time-dependent current. We find that the inter-band and intra-band contributions display very different time-frequency characteristics. We show that solutions in these two bases are equivalent under an unitary transformation but that, unlike the velocity gauge method, 
the Houston state treatment is numerically unstable when more than a few low lying energy bands are used.

\end{abstract}

\maketitle

\section{Introduction}
Since high harmonic generation (HHG) in inert gases was first discovered in 1987 \cite{McPherson1987, Ferray1988}, it has become one of the major research areas in ultrafast atomic physics. In three decades of development, HHG has pushed the technology for creating tunable extreme ultraviolet (XUV) pulses into the attosecond regime \cite{Paul2001, Hentschel2001, Krausz2009}, and has been widely used to probe the ultrafast dynamics of atomic and molecular, and solid systems \cite{Uiberacker2007, Kelkensberg2009, Schiffrin2013, Leone2014}. Since the HHG process is highly nonlinear, the intensity of the generated harmonics is typically orders of magnitude lower than the driving laser intensity. This limits the number of photons per pulse that can be obtained, meaning that  applications such as pump-probe spectroscopy and lithography using HHG are not presently practical using gas-phase sources.

Recently, Ghimire \textit{et al.} discovered that high order harmonics can also be generated from a bulk crystal \cite{Ghimire2010}, which has opened up the possibility of studying attosecond electron dynamics in materials. Because of the use of a high-density target, solid-state HHG has a potential for high efficiency. In addition, it may be possible to engineer the structure of the solid target on a micrometer scale, and thereby design periodic structures that enhance the macroscopic phase matching \cite{Gibson2003,Zhang2007,Zepf2007,Seres2007,Tosa2008}, further boosting the number of photons generated. From a fundamental point of view, solid-state HHG is also interesting as a potential tool for addressing and understanding the ultrafast dynamics of electrons in periodic structures.

\begin{figure}[h]
\centering
\includegraphics[width=0.35\textwidth]{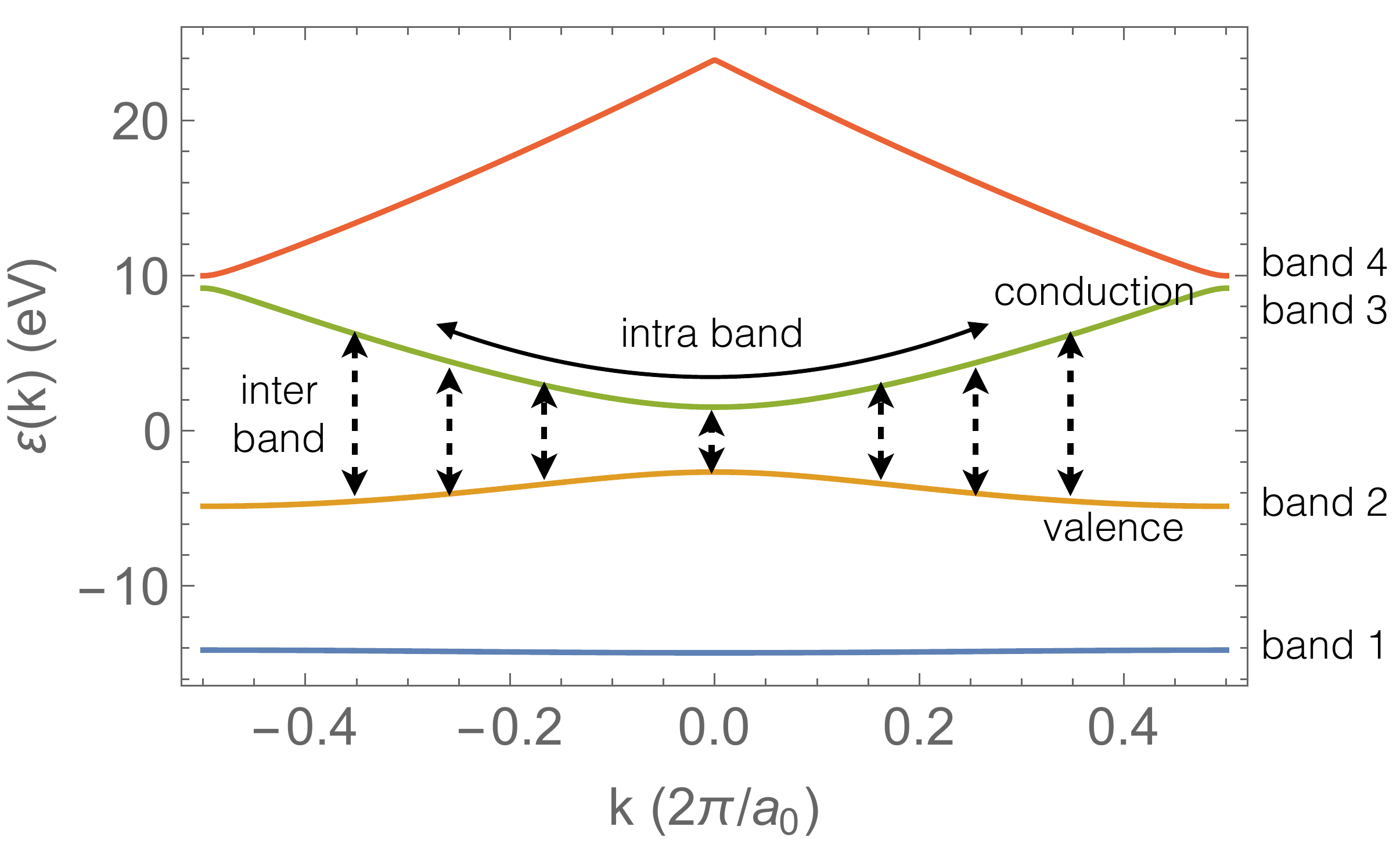}
\caption{(Color online) The band structure used in our calculation and the scheme of the inter-band and intra-band dynamics. The intra-band dynamics involves the motion of the electron on the same band, while inter-band dynamics describes the transitions of the electron between different bands. 
We regard the second band as the valence band and the third band as the conduction band.} 
\label{fig:band_structure}
\end{figure}

The electron dynamics in a solid interacting with an electromagnetic field are generally considered to have a contribution from both intra-band and inter-band dynamical processes \cite{Golde2008,Rossi1998, Vampa2014}, as illustrated in Fig.~\ref{fig:band_structure}.  The intra-band dynamics involves $k$-space motion of an electron along one (or several) specific bands, while the inter-band dynamics involves  electron transitions between different bands.  Although this picture has been widely adopted in studying dynamics in solids, the mechanism for solid-state HHG has still not been well characterized in these terms. Ghimire {\it et al.} \cite{Ghimire2010,Ghimire2012} proposed that laser-driven Bloch oscillations of an electron wave packet on a single conduction band could be the source of the non-linear current responsible for HHG. This model was supported by recent results on THz HHG by Schubert {\it et al.} \cite{Schubert2014}. In contrast, calculations by Vampa {\it et al.} \cite{Vampa2014} using a two-band model indicated that the HHG spectrum is dominated by the inter-band current and furthermore that the cutoff is limited by the largest band gap. Hawkins {\it et al.} \cite{Hawkins2015} suggested that higher-lying bands should be included in the laser-solid description in order to accurately capture the laser-driven electron dynamics. 

A theoretical framework for the interaction of lasers with solids has been constructed using both many-electron models \cite{Schubert2014, Meier1994,Golde2008, Kemper2013} and single-electron models \cite{Faisal1997a,Apalkov2012, Vampa2014, Hawkins2015, Higuchi2014a,Mucke2011}. The many-electron models are based on second quantization, together with a reduction in correlation via a Hartree-Fock decoupling scheme, which leads to the well-known semiconductor Bloch-Equations (SBE) \cite{Simon1993}. In the simplest case, the SBE describe an ensemble of correlated two-level systems \cite{Klingshirn2007}. The SBE approach has been used extensively in describing optical properties of solids, where it successfully describes many semiconductor optical experiments such as pump-probe, four-wave-mixing, and photon echoes \cite{Klingshirn2007, Haug2004, Shah1996a,Schafer2002a}. The single-electron models, on the other hand, treat the solid as a single electron in an effective periodic potential. The laser-solid interaction is then described by the laser-driven single-electron motion in this effective periodic potential \cite{Wannier1960}. This single-electron approach has been very successful in addressing electron dynamics in periodic structures such as Bloch oscillations, Zener tunneling, and Wannier-Stark localization in semiconductor superlattices \cite{Mendez2008,Gluck2002}, optical lattices \cite{Dreisow2009,Trompeter2006} as well as waveguide arrays \cite{Block2014,Longhi2006}. We note that the single-electron models are conceptually the closest to the hugely successful single-active-electron treatment of HHG in atomic and molecular gases, which has yielded a number of insights into both the mechanism and control over the harmonic and attosecond pulse generation processes \cite{Schafer1993, Corkum1993, Lewenstein1994, Hentschel2001, Paul2001, Agostini2004, Lin2010}.

In this paper, we will follow the analogy to optical lattices and HHG in gases and work in a single-electron framework. We solve the time-dependent Schr\"odinger equation (TDSE) for an electron interacting with a periodic potential using two different numerical approaches, which allows us a number of insights into the HHG process. In the first approach, we follow the velocity gauge treatment of Korbman {\it et al.} \cite{Korbman2013} in which the wave function is expanded in a basis of Bloch states, which means that many bands are included in the calculation. We find that this method is numerically stable with respect to increasing both the laser intensity and the number of bands considered, but that it does not allow us to separately consider the intra-band and inter-band electron dynamics. Our second approach is to solve the TDSE in a time-dependent basis set \cite{Krieger1986}, the so-called Houston states, in which the intra- and inter-band contributions can be naturally separated. We show that while the two methods are equivalent under a unitary transformation, the Houston state treatment becomes numerically unstable as the number of bands is increased. 

We find that the resulting harmonic spectra exhibit both a primary and a secondary plateau, each with a cutoff energy that depends linearly on the laser electric field strength. The primary plateau is dominated by  inter-band transitions between the valence band and the first conduction band. The secondary plateau is due to transitions between the valence band and the higher-lying conductions bands. This plateau is much weaker than the primary plateau at low intensity, but increases rapidly with the intensity and eventually merges with the primary plateau. 
Using the Houston state approach, we also separately analyze the time-frequency characteristics of the inter- and intra-band contributions to the current and find that they exhibit very different characteristics, in the intensity regime where the spectrum is dominated by one primary plateau. We propose that this difference could potentially be used to experimentally address which mechanism dominates the solid-state HHG process.

This paper is organized as follows: In Section II, we present the theoretical framework and the results of solving the TDSE using the velocity gauge approach, and in Section III, we solve the TDSE in the Houston basis \cite{Krieger1986} and analyze the results of section II in terms of  inter-band and intra-band dynamics. Section IV presents an analysis of the numerical instability of the Houston basis treatment, and Section V presents a brief summary.

\section{TDSE in the Bloch State Basis}

We consider a linearly polarized laser field propagating through a thin crystal along the optical axis. We describe the laser-solid interaction in one dimension, along the laser polarization which lies in the crystal plane. We follow the velocity gauge treatment in \cite{Korbman2013}, in which the TDSE reads
\begin{equation}
\label{Eq:TDSE_velocity}
i\hbar\frac{\pd}{\pd t}\ket{\psi(t)}=(\hat H_0+\hat H_\text{int})\ket{\psi(t)},
\end{equation}
where $H_0$ is the field-free Hamiltonian and $H_\text{int}$ is the interaction Hamiltonian between the laser and the electron
\begin{align}
\hat{H}_0 &= \frac{\hat{ p}^2}{2m} + V(\hat{x})\\
\hat{H}_\text{int} &= \frac{e}{m}A(t)\hat{ p}.
\end{align}
$A(t)$ is the vector potential, and is related to the electric field by
\begin{equation}
\label{eqn:define_A(t)}
A(t)=-\int_{-\infty}^t E(t') dt'.
\end{equation}
 $\hat{p}$ is the momentum operator and $V(x)$ is the periodic lattice potential. We have employed the dipole approximation $A(x,t)\approx A(t)$ because the wavelengths we are interested in are much larger than the lattice constant. According to Bloch's theorem, the eigenstates of the field-free Hamiltonian are the Bloch states
\begin{align}
\hat{H}_0\ket{\phi_{nk}} &= \varepsilon_{n}(k)\ket{\phi_{nk}},
\end{align}
where $n$ is the band index and the eigenvalues $\varepsilon_{n}(k)$ represent the dispersion relations of the bands. Each Bloch state can be written as a product of a plane wave and a function periodic in the lattice spacing $a_0$:
\begin{equation}
\braket{x|\phi_{nk}} = e^{ikx}U_{nk}(x),
\end{equation}
where $U_{nk}(x)$ satisfy
\begin{equation}
U_{nk}(x+a_0)=U_{nk}(x).
\end{equation}
Because the vector potential is independent of $x$, the lattice momentum $k$ is still a good quantum number \cite{Holthaus1993}. This means  the dynamics of the different lattice momentum channels are independent, and the TDSE can be solved independently for each $k$ \cite{Korbman2013}.

To solve the TDSE for a specific $k_0$, we express the wave function in Bloch states
\begin{align}
\ket{\psi_{k_0}(t)} &= \sum_n C_{nk_0}(t) \ket{\phi_{nk_0}},
\label{eqn:expand_wave_function_in_Bloch_states}
\end{align}
and solve for the time-dependent coefficients $C_{nk_0}(t)$
\begin{equation}
i\hbar\frac{\partial}{\partial t} C_{nk_0} = C_{nk_0} \varepsilon_n(k_0) + \frac{e A}{m}\sum_{n'} C_{n'k_0} p^{nn'}_{k_0},
\end{equation}
where the $p_{nn'}$ matrix element is the integration of the momentum operator over a lattice cell in space
\begin{align}
p^{nn'}_{k_0}&=\braket{\phi_{nk_0}|\hat p|\phi_{n'k_0}}\notag\\
&=\frac{1}{a_0}\int_0^{a_0} dx\, \phi_{nk_0}^*(x) \left(\frac{\hbar}{i}\frac{\partial}{\partial x}\right)\phi_{n'k_0}(x).
\end{align}
Usually, the $p$ matrix is dominated by its tri-diagonal matrix elements, which means the transitions to higher bands are most likely to happen through successive transitions between intermediate bands. Finally, we calculate the time-dependent laser-induced current as the sum of of the current in each of  the different $k_0$ channels $j_{k_0}$ \cite{Korbman2013} where :
\begin{equation}
 j_{k_0}  = -\frac{e}{m} \left[\text{Re}\left[\braket{\psi_{k_0}|\hat p|\psi_{k_0}}\right]+eA(t)\right].
\label{Eq:total_current_Bloch_basis}
\end{equation}
%

The laser pulse we use has a $\cos^4$ envelope in its electric field, with a full width at half maximum (FWHM) pulse duration of 3 optical cycles for all  wavelengths. We have considered laser wavelengths $\lambda$ between 2~$\mu$m and 5~$\mu$m, and intensities between 1$\times 10^{10}$ W/cm$^2$ and 2$\times 10^{12}$ W/cm$^2$. The harmonic spectrum is calculated as the modulus square $|j(\omega)|^2$ of the Fourier transform of the time-dependent current in Eq.(\ref{Eq:total_current_Bloch_basis}). Before the Fourier transform, we multiply $j(t)$ by a time-dependent window function  in order to suppress the coherence between population in the conduction and valence bands which would otherwise last forever (in our model) and would dominate the spectrum in the region  around the band-gap energy. The window function matches the envelope of the laser pulse.

Throughout the paper, we use a periodic potential $V(x)=-V_0(1+\cos(2\pi x/a_0))$, with $V_0 = 0.37$ and lattice constant $a_0=8$, both in atomic units. This Mathieu-type potential leads to a band structure that can be expressed in terms of Mathieu functions \cite{Slater1952}. It has been used extensively in the optical lattice community \cite{Hartmann2004a,Breid2006, Chang2014}. The resulting band structure has a minimum band gap of 4.2 eV and is shown in Fig.~\ref{fig:band_structure}. Unless otherwise specified, we have used 51 Bloch states in our expansion of the wave function, which means that 51 bands are included in the calculations for each $k$ value. Since the lowest band (band 1) is deeply bound and very flat, we use band 2 as the initially populated valence band. We have checked that transitions involving band 1 play a negligible role in the harmonic generation dynamics. The initial population is  a small superposition ($\Delta k_0=\pi/20 a_0$) of Bloch states near $k_0=0$ on the valence band, corresponding to a  wave function which is initially spatially delocalized  through-out the solid. This implies that the valence band is near ``frozen" so that only a small distribution of population near $k=0$ can be excited to higher bands,  and is similar to the initial condition proposed in \cite{Ghimire2012}, where only a small part (about 2\% in our case) of valence band electrons are excited to the conduction band where they undergo laser-driven Bloch oscillations. It is also similar to  the initial condition used in quantum well and optical lattice systems when inducing Bloch oscillations \cite{Simon1993, Wannier1962, Breid2006,Hartmann2004a, Chang2014}. We note that Bloch oscillations (usually thought of as electron motion in $k$-space) can indeed be captured  by our formalism in which different $k$'s are uncoupled from each other. In this formalism, Bloch oscillations are more easily conceptualized as charge oscillations in real space, rather than in $k$ space \cite{Breid2007}.

\begin{figure}[h]
\centering
\includegraphics[width=0.4\textwidth]{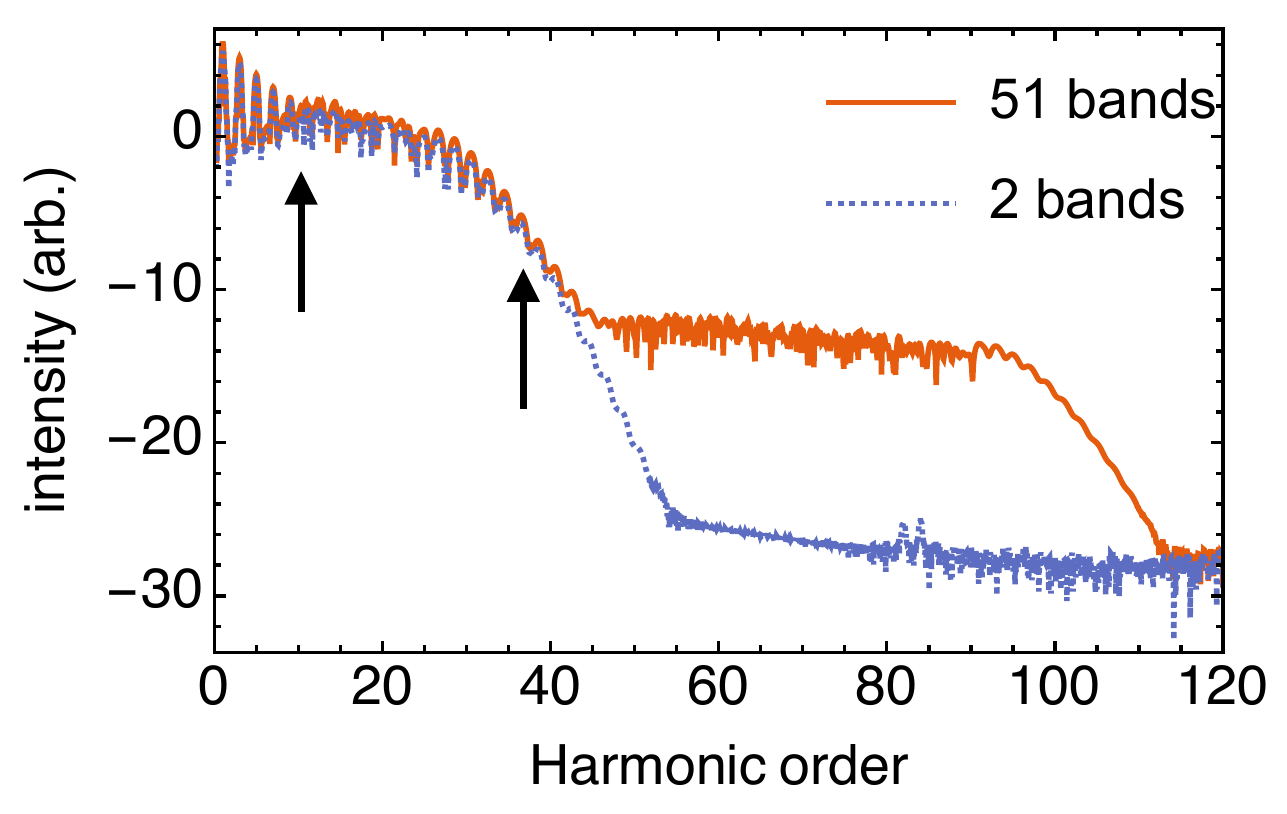}
\caption{(Color online) High harmonic spectra (logarithmic intensity scale) of the laser induced current calculated by solving the TDSE in velocity gauge. The laser wavelength and peak intensity are 3.2~$\mu$m and $4.5\times 10^{11}$ W/cm$^2$. The red solid curve shows the spectrum where 51 bands are used in the calculation whereas the blue dashed curve uses the same condition as the red curve, except  bands 4 and 5 are removed from the calculation. The black arrow indicates the minimum and maximum band gap energies.} 
\label{fig:Multiband_spectrum}
\end{figure}

\begin{figure}[h]
\centering
\includegraphics[width=0.48\textwidth]{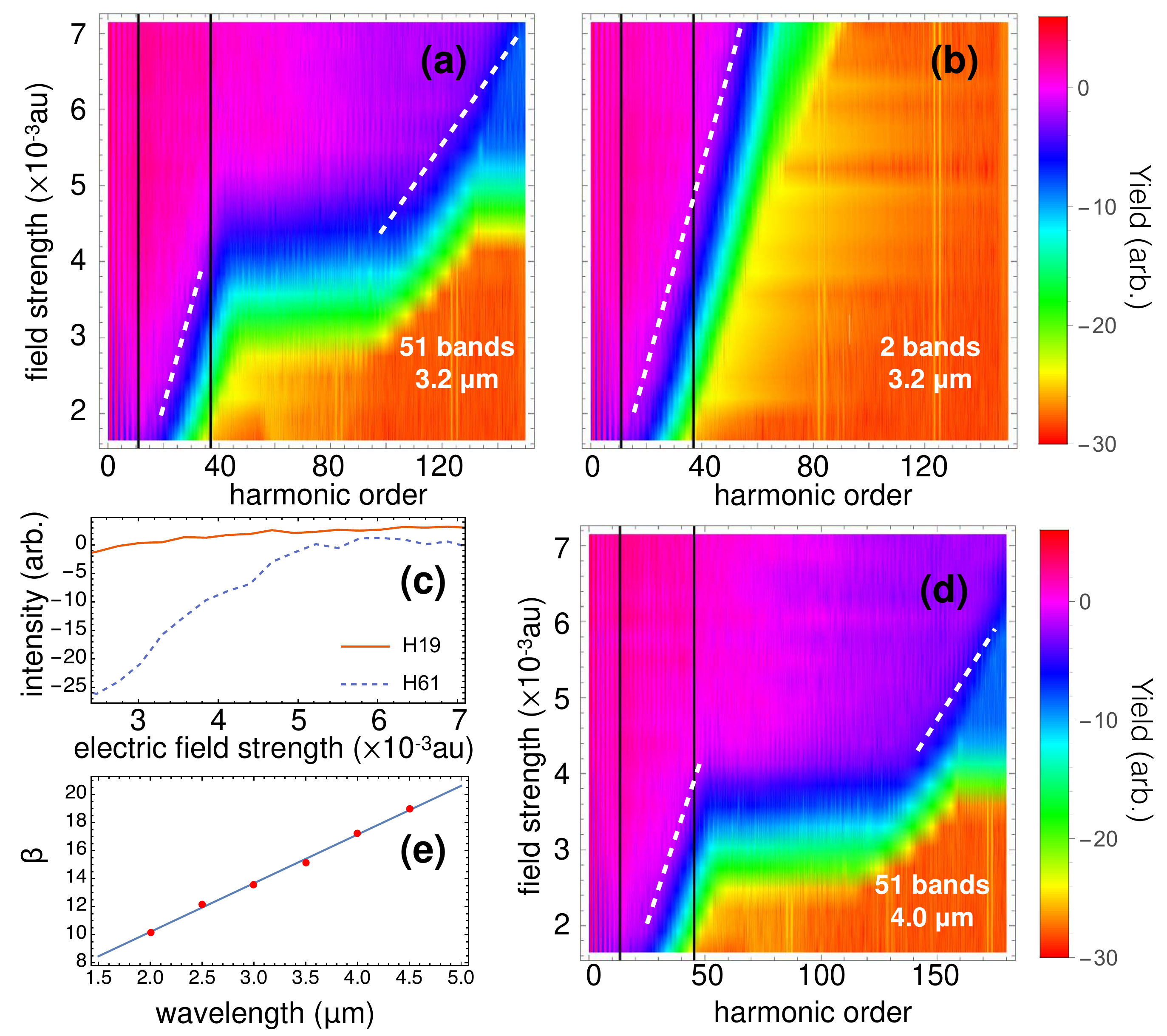}
\caption{(Color online) (a) Harmonic yield as a function of laser field strength for $\lambda = 3.2$~$\mu$m. (b) Same as (a) but excluding bands 4 and 5. White dashed lines indicate the linear dependence of the cutoff energy on field strength, and vertical black lines indicate the minimum and  maximum band gaps between the valence and conduction bands. (c) Field strength dependence of H19 and H61 from (a). (d) Same as (a) but using $\lambda = 4.0$~$\mu$m. (e) Wavelength dependence of the cutoff scaling coefficient with electric field strength. All yields are shown in logarithmic scale.} 
\label{fig:Multiband_spectrum_intensity_scan}
\end{figure}

Fig.~\ref{fig:Multiband_spectrum} shows harmonic spectra for our model system calculated using a laser wavelength $\lambda = 3.2$~$\mu$m and intensity $4.5\times 10^{11}$ W/cm$^2$. This corresponds to a Bloch frequency $\omega_B=Ea_0/\hbar=0.78$ eV, where $E$ is the electric field amplitude. Although this intensity is low compared to the experiment in \cite{Ghimire2010}, it is high enough to generate rich nonlinear dynamics. The harmonic spectrum exhibits both a perturbative regime (harmonic order $<10$), a plateau regime ($10-30$) and a cutoff ($\sim 30$), very similar to the general structure of the harmonic spectrum generated by atoms \cite{Schafer1993,Corkum1993}. As we will show in the following section, the plateau is due to inter-band transitions between the conduction and valence band. This agrees with the prediction in \cite{Vampa2014}. However, in contrast to that paper, we find that harmonics can be generated with photon energies well above the minimum and maximum band gap energies, as shown in Figs.~\ref{fig:Multiband_spectrum} and \ref{fig:Multiband_spectrum_intensity_scan}.  

The two curves in Fig.~\ref{fig:Multiband_spectrum} represent the full calculation, including all 51 bands, and a calculation in which bands 4 and 5 (the second and third conduction band) have been dynamically excluded. This is done by setting the coefficients of $C_{4k_0}$ and $C_{5k_0}$ to zero in Eq.~\eqref{eqn:expand_wave_function_in_Bloch_states} at each time step in solving the TDSE. We note that since the transition probability from bands 2 and 3 to  bands 6 and above is very small, removing bands 4 and 5 makes our model effectively a two-band model. The full-calculation spectrum exhibits a weak second plateau, about 10 orders of magnitude lower than the main plateau, which is absent in the reduced-band calculation. The comparison between the two curves yields two insights: (i) the main part of the plateau and cutoff region is well described by the dynamics involving just the valence and the lowest conduction bands, indicating that higher bands play a negligible role in the harmonic generation in this wavelength and intensity regime, and (ii) the second plateau is due to contributions to the dynamics involving  higher-lying bands,  predominantly due to transitions between bands 4 or 5 and the valence band. 

We next investigate the intensity and wavelength dependence of the first and second plateau and cutoff. 
Fig.~\ref{fig:Multiband_spectrum_intensity_scan}(a) shows the harmonic yield as a function of laser electric field strength for $\lambda = 3.2$~$\mu$m. We  see that the cutoff of the first plateau increases linearly with field strength. This is in agreement with the experimental finding in \cite{Ghimire2010}. Fig.~\ref{fig:Multiband_spectrum_intensity_scan}(b) shows the same intensity scan without bands 4 and 5, and the linearity of the first cutoff is revealed for a larger range. Going back to Fig.~ \ref{fig:Multiband_spectrum_intensity_scan}(a), we see that as the intensity increases, the second plateau rises and merges with the first plateau consistent with what was observed in \cite{Faisal1996}. Subsequently, the cutoff energy of this new, longer plateau,  also exhibits a linear dependence on the laser field strength. 
Fig.~\ref{fig:Multiband_spectrum_intensity_scan}(c) explores the build-up of the second plateau in more detail. We show a line-out of the field-strength dependence of harmonics 19 and 61 (H19 and H61), which are located in the first and second plateau respectively. At the highest fields, these harmonics are both in the plateau and therefore have similar yields and change only slowly with field strength. However, at the lowest field strengths, H61 is essentially zero (not shown in  Fig.~\ref{fig:Multiband_spectrum_intensity_scan}(c)) until it starts to increase exponentially with intensity, indicating that the second plateau is not independent but rather built off of the first plateau. This is consistent with our finding above, that the population in the higher bands is built in a step-like process based on first populating lower bands. 

Our model predicts similar harmonic generation dynamics at other wavelengths. Fig.~\ref{fig:Multiband_spectrum_intensity_scan}(d) shows the field strength dependence of the harmonic spectrum at $\lambda = 4.0$~$\mu$m, which also exhibits two plateau regions with two different, linear, dependences of the cutoff energy on the field strength. 

From our numerical results, we can quantify the scaling of the cutoff energy in the following way. We start by writing the cutoff energy ${\cal E}_{\text{cutoff}}$ in units of the Bloch frequency, as:
\begin{equation}
{\cal E}_{\text{cutoff}} \approx \beta \omega_B ,
\end{equation}
then the scaling factor for the first and second plateaus are $\beta_1=14$, $\beta_2=38$, respectively. Fig.~\ref{fig:Multiband_spectrum_intensity_scan}(e) shows the scaling factor for 6 different wavelengths for the first plateau and suggests that the cutoff also scales approximately linearly with the wavelength. Thus, in this model, the first cutoff energy depends linearly on both the electric field strength and the wavelength: 
\begin{equation}
{\cal E}_{\text{cutoff}} \propto \lambda E ,
\end{equation}
where the proportionality constant depends on the band structure. The linear dependence on field strength contrasts with the $(\lambda E)^2$ scaling of the cutoff in atomic and molecular gases \cite{Krause1992}, but agrees well with the prediction for a strongly driven two-level system \cite{Gauthey1997, Krainov1994} which in our case would be ${\cal E}_\text{cutoff} = p_{vc}E\lambda/\pi c$, where $p_{vc}$ is the momentum operator matrix element between the valence and conduction bands at $k=0$. The two-level formula underestimates our numerical values for $\beta$ by about 10\%. The scaling of the second cutoff with wavelength is more difficult to quantify. Both Fig.~\ref{fig:Multiband_spectrum_intensity_scan}(a) and (d) suggest that a third plateau appears at the highest energies, possibly due to the contribution of bands 6 and 7. Apart from concerns about applying our simple model to such high intensities, we note that there are as of yet no high harmonic experiments with the level of sensitivity that would be needed to observe such an effect.

To conclude this section, we briefly comment on our choice of initial condition in which a small \mbox{$k$-state} wave packet yields an initial wave function which is spatially delocalized across the entire (1D) crystal. Other recent calculations have considered a different initial condition in which the valence band is initially fully populated \cite{Hawkins2015,Vampa2014}, which in our model would correspond to an initial wave function localized at one particular lattice site. In a real insulating material, the filled valence band  means that all the different electronic states of the valence band are occupied by different electrons. The {\it full} valence band thus only has meaning in the multi-electron context. In a single-electron framework the valence band can never be filled in the same way, since we only have one electron, which corresponds to a much lower dimensional Hilbert space than the multi-electron wave function. In this sense, in the single-electron framework the solid is modeled like a super-atom with a atomic potential that is periodic. What we can choose is only the initial wave function for this super-atom. For instance, we can choose the initial condition of the super-atom to be a Bloch state (few $k$'s, spatially delocalized) or a Wannier state (many $k$'s, spatially localized). As demonstrated recently in rare-gas clusters \cite{Park2014}, the delocalization of the initial wave packet may  affect the HHG process, which could also be the case in solids.

\section{Houston State Basis}
In the previous section, the electron dynamics was described in a static basis of Bloch states using the velocity gauge interaction. In this picture  the time dependence of the wave function is due solely to the time dependence of the Bloch state coefficients $C_{nk}(t)$. Though computationally convenient, this 
method provides a time-dependent current which is hard to understand at an intuitive level. For example, the familiar Bloch oscillation of an electron with a momentum $k_0$ in a static field is built from the superposition of a large number of bands all at the same $k_0$. Obviously, in this picture there can be no separation of the current into intra-band and inter-band contributions.

In this section we describe an alternative way to calculate the electron dynamics using a time-dependent basis set, the Houston states \cite{Houston1940}. As demonstrated in Appendix A, the two solutions are equivalent, since they are related by a unitary frame transformation. In the Houston basis, however, we can obtain a separation of the induced current into intra- and inter-band components. This will allow us to separately explore the time-frequency characteristics of the two contributions, and show that they exhibit very different emission times.

The Houston  states are best thought of as an adiabatic basis in which the lattice momentum that would be $k_0$ in the absence of a field has the time-dependence:
\begin{equation}
k(t)=k_0+\frac{eA(t)}{\hbar}.
\label{eqn:lattice_momentum_relationship_A}
\end{equation}
By construction they are the instantaneous eigenstates of the time-dependent Hamiltonian $H(t)$:
\begin{align}
H(t) \ket{\widetilde\phi_{nk_0}(t)} = {\varepsilon}_{n}(k(t)) \ket{\widetilde\phi_{nk_0}(t)},
\label{eqn:Houston_state_as_adabatic_state}
\end{align}
where $H(t)$ is the Hamiltonian in the same single-electron Schr\"odinger equation as above Eq.~\eqref{Eq:TDSE_velocity} \cite{Krieger1986} except for an additional term proportional to  $A^2$:
\begin{equation}
i\hbar\frac{\pd}{\pd t}\ket{\psi(t)} = \left[\frac{\left(\hat{ p} + e A\right)^2}{2m} +V(x)\right]\ket{\psi(t)}.
\label{eqn:Houston_TDSE}
\end{equation}
Including this term in the Schr\"odinger equation makes the form of the Houston states simpler, but it has no effects on the current since the wave function only differs by an overall time-dependent phase. In this convention, the Houston states are related to the Bloch states with lattice momentum $k(t)$ by \cite{Krieger1986}
\begin{equation}
\ket{\widetilde\phi_{nk_0}(t)} = e^{-i e A\hat x/\hbar}\ket{\phi_{nk(t)}},
\end{equation}
where $\hat{x}$ is the position operator. Expanding the time dependent wave function with initial lattice momentum $k_0$ in Houston states
\begin{equation}
\ket{\psi(t)} = \sum_n a_{nk_0}(t) \ket{\widetilde\phi_{nk_0}(t)},
\label{eqn:wavefunction_expand_in_Houston_basis}
\end{equation}
we find equations of motion for the coefficients
\begin{equation}
i\hbar \frac{\partial a_{nk_0}}{\partial t} = \sum_{n'}\Big[\varepsilon_n(k(t)) \delta_{nn'} - eE(t) X_{nn'}(k(t)) \Big]a_{n'k_0}.
\label{eqn:EOM_Houston}
\end{equation}
where we have made use of Eq.~(\ref{eqn:define_A(t)}). $X_{nn'}$ is the inter-band transition matrix element defined by
\begin{equation}
X_{nn'}(k)=\frac{1}{ia_0}\int_0^{a_0} U^*_{nk}\frac{\pd}{\pd k} U_{n'k} dx
\label{eqn:X_matrix}
\end{equation}
where $X_{nn'}=0$ if $n=n'$. It is calculated numerically using procedures in \cite{Lindefelt2004}. Note that the initial wave function has a lattice momentum of $k_0$, but we could also make it a wave packet as we did in the previous section. 

The time-dependent Houston states describe the electron dynamics in a moving frame in which the lattice momentum is prescribed by the vector potential as in Eq.~\eqref{eqn:lattice_momentum_relationship_A}. Pictured in $k$ space, one can think of an electron wave packet oscillating on each energy band, while at the same time some of the amplitude transitioning between different bands, corresponding to the intra- and inter-band dynamics, respectively. The motion on each band is governed by the time-dependent dispersion $\varepsilon_n(k(t))$. For the intensity used in Fig.~\ref{fig:Multiband_spectrum}, the motion of the wave packet in $k$ space samples about $2/3$ of the first Brillouin zone. 

The total current can be calculated from Eq.~\eqref{Eq:total_current_Bloch_basis}, using  Eq.~\eqref{eqn:wavefunction_expand_in_Houston_basis} for the wave function: 
\begin{equation}
j_\text{tot}  = -\frac{e}{m} \text{Re}\left[ \sum_{nn'} a^*_{nk_0} a_{n'k_0}\braket{\phi_{nk(t)}|\hat p|\phi_{n'k(t)}} \right].
\label{eqn:Houston_total_current}
\end{equation}
Since now the system is described in the frame that moves along with the field, there is no $A$ term in the expression for the current, as opposed to that of Eq.~\eqref{Eq:total_current_Bloch_basis}. 

\begin{figure}[h]
\centering
\includegraphics[width=0.4\textwidth]{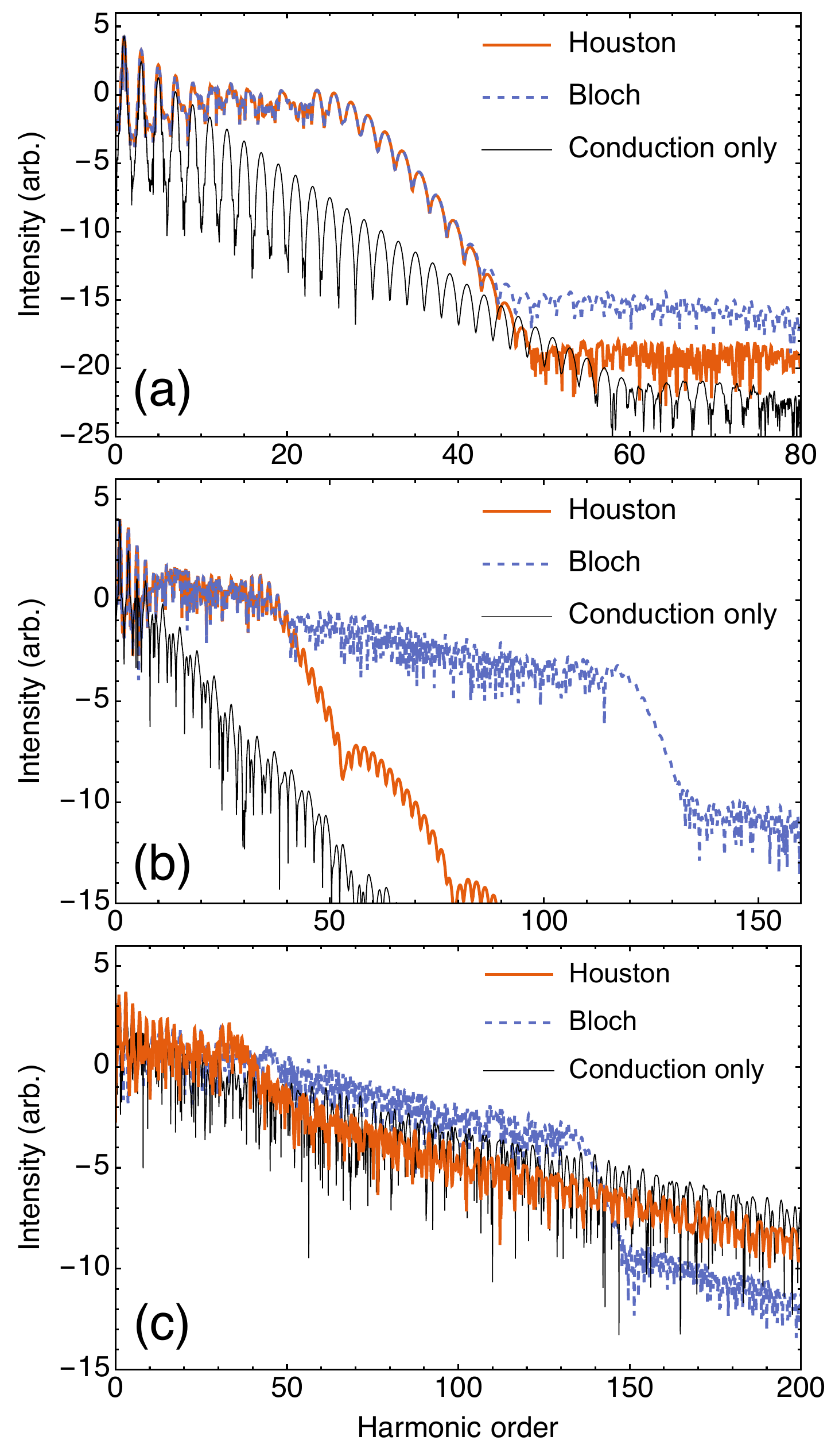}
\caption{(Color online) Comparison of harmonic spectra (logarithmic scale) from a three-band Houston basis calculation and a 51-band Bloch basis calculation, for three different intensities of a 3.2~$\mu$m driving field, (a) $4.5 \times 10^{11}$~W/cm$^2$, (b) $1.0 \times 10^{12}$~W/cm$^2$, and (c) $1.8 \times 10^{12}$~W/cm$^2$. The initial condition used in both calculations is a single $k=0$ in the valence band. The thin line in (a) and (c) is calculated from the conduction band only, in the Houston basis, by ignoring inter-band transitions (see text). } 
\label{fig:Multiband_VS_Houston_spectrum}
\end{figure}

Fig.~\ref{fig:Multiband_VS_Houston_spectrum} compares spectra calculated in the Houston basis and the Bloch basis at three different field strengths, corresponding to the electron wave packet in $k$-space sampling about 2/3, 3/3, and 4/3 of the first Brillouin zone.  In the Houston basis, we use the three lowest bands shown in Fig.~\ref{fig:band_structure}, while in the Bloch state basis we use as many of the 51 bands that were used to calculate the band structure  as are necessary for numerical convergence. Of these three bands, only the valence and conduction bands (2 and 3) contribute meaningfully to the dynamics. The reason for using only three bands in the Houston basis will be discussed below in connection with the numerical properties of the $X$ matrix elements. The initial condition used in both cases is a single $k$ state ($k=0$) in the valence band where the band gap is the smallest. Fig.~\ref{fig:Multiband_VS_Houston_spectrum}(a) and (b) shows that the agreement between the Houston and the Bloch calculations is excellent for the part of the spectrum that is dominated by the dynamics in the valence and conduction band only. This agreement is expected, since the two wave functions are related by a unitary transformation (see Appendix A). Since the Houston calculation only includes three bands, it cannot be expected to reproduce the high-frequency part of the spectrum that in the Bloch calculation is due to higher-lying bands (approximately above harmonic order 50 in both Fig.~\ref{fig:Multiband_VS_Houston_spectrum}(b) and (c)). It is worth noting, though, that at the highest intensity the slope of the Houston spectrum matches that of the Bloch spectrum. We will comment on this in more detail below.

\begin{figure}[h]
\centering
\includegraphics[width=0.4\textwidth]{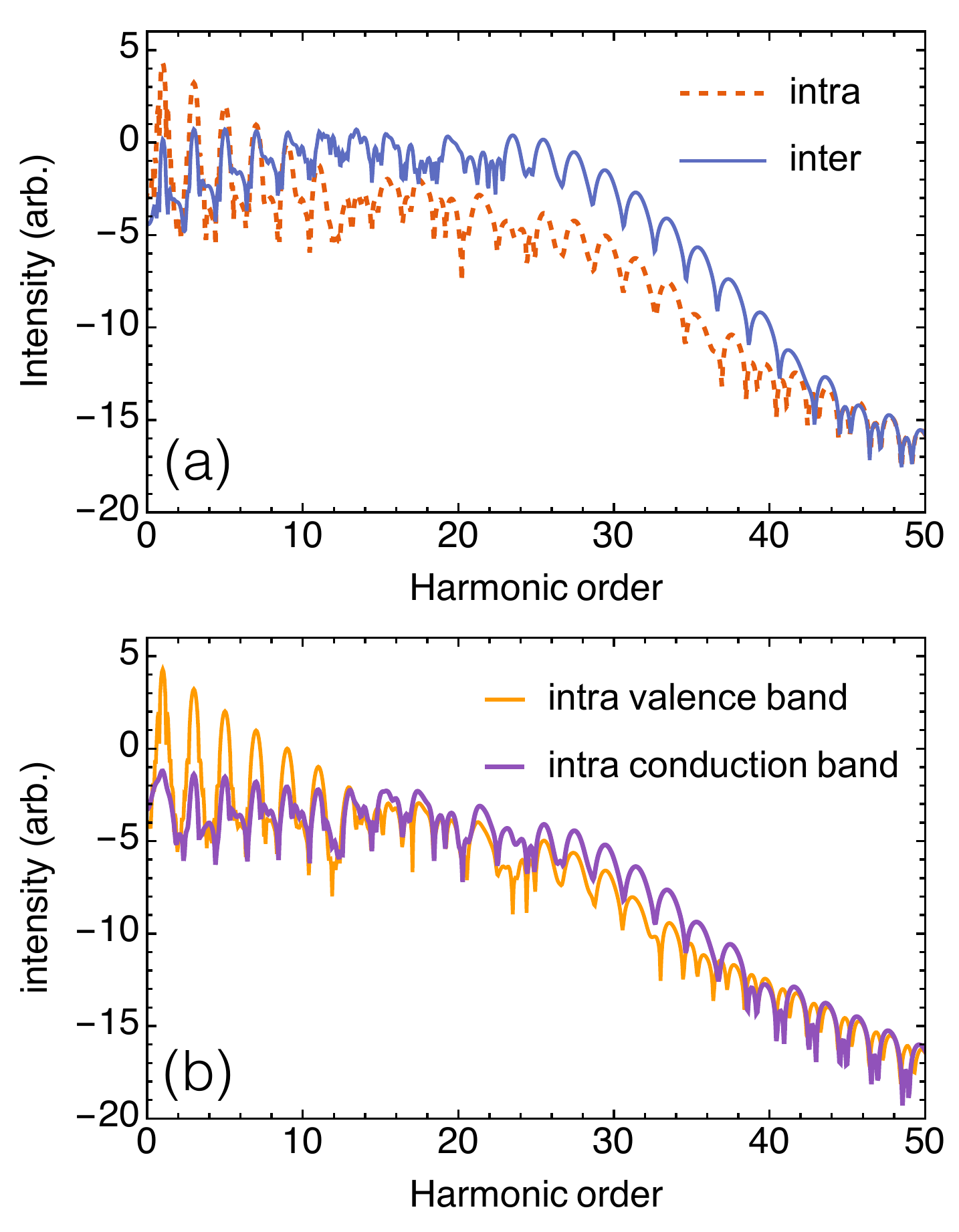}
\caption{(Color online) (a) The intra-band (red, dashed) and inter-band (blue, solid) current (logarithmic scale) of a Houston model. The low order harmonics mainly come from intra-band current while the harmonics in the plateau are mainly come from the inter-band current. (b) The valence band (orange) and conduction band (purple) contributions to the intra current (logarithmic scale).} 
\label{fig:Houston_intra_inter_spectrum}
\end{figure}

One advantage of the Houston basis is that the electron dynamics naturally separate into an intra- and inter-band contribution, and can be studied separately. In Eq.~\eqref{eqn:Houston_total_current}, the intra-band contribution to the current involves only Houston states on the same band ($n=n'$), whereas the inter-band contribution involves transitions between different bands ($n\neq n'$).
\begin{eqnarray}
j_\text{intra}   &=& -\frac{e}{m}  \sum_{n} \left| a_{nk_0}\right|^2 \braket{\phi_{nk(t)}|\hat p|\phi_{nk(t)}}\label{eqn:Houston_current_intra}\\
j_\text{inter}   &=& -\frac{e}{m}  \text{Re}\left[ \sum_{\substack{n,n' \\ n \neq n'}} a^*_{nk_0} a_{n'k_0}\braket{\phi_{nk(t)}|\hat p|\phi_{n'k(t)}} \right]\notag\\
\label{eqn:Houston_current}
\end{eqnarray}
The intra- and inter-band contributions to the current are shown in Fig.~\ref{fig:Houston_intra_inter_spectrum}(a), for the Houston spectrum shown in Fig.~\ref{fig:Multiband_VS_Houston_spectrum}(a). We find that for the range of intensities where the three-band Houston model is applicable, the inter-band contribution to the plateau in the  spectrum is stronger than the intra-band contributions by several orders of magnitude in the plateau regime. This is in agreement with the prediction in \cite{Vampa2014}. In Fig.~\ref{fig:Houston_intra_inter_spectrum}(b), we show the intra-band contribution from the valence and conduction band separately. This is done by plotting separately the terms in the sum in Eq.~\eqref{eqn:Houston_current_intra}. We note that the intra-band contribution from the valence band would be unphysical for a real insulator in which the valence band would be filled. In the Bloch state basis, we have performed calculations in which we used a full valence band as the initial condition. This suppresses the yield of the few lowest harmonics but otherwise leads to a harmonic spectrum that agrees well with those in Fig.~\ref{fig:Multiband_VS_Houston_spectrum}(a) until approximately harmonic 30. We conclude from this that the (overestimated) contribution from driven Bloch oscillations in the valence band can be ignored for the range of frequencies we are interested in. 

It is interesting to note that  the single conduction band model used in \cite{Ghimire2012,Ghimire2010} comes naturally from the Houston model if one starts with the initial population in the conduction band and eliminates  inter-band transitions. When the inter-band transition matrix $X$  vanishes, the total current reduces to the intra-band current expression used in \cite{Ghimire2012}, where the current originates in the electron motion in the conduction band (see proof in Appendix B). The resulting spectrum is plotted  in Fig.~\ref{fig:Multiband_VS_Houston_spectrum} (labeled "conduction only") , normalized to the response at the fundamental. For the lower intensities, the intra-conduction-band current spectrum is completely different than the full (Bloch basis) spectrum and does not exhibit a plateau or cutoff. This is in agreement with the finding that the harmonic spectrum is dominated by inter-band transitions. It also shows that the clear cutoff that can be seen in the conduction band spectrum in Fig.~\ref{fig:Houston_intra_inter_spectrum}(b) is in fact due to the time-dependence of the population transfer in and out of the conduction band (through the time-dependence of the transition matrix X), rather than due to the intra-band dynamics itself. At the highest intensity, Fig.~\ref{fig:Multiband_VS_Houston_spectrum}(c), the slope of the intra-conduction-band spectrum agrees very well with that of the full calculation, but is again lacking a cutoff energy. In Section II, we interpreted the extended (secondary) plateau in the harmonic spectrum as being due to transitions involving high-lying bands, when describing the dynamics in the Bloch state picture. The result in Fig.~\ref{fig:Multiband_VS_Houston_spectrum}(c) suggests that an alternative interpretation is that the extended plateau is due to driven conduction-band Bloch oscillations traversing the entire Brillouin zone, but that in this model the cut-off energy cannot be captured without considering inter-band transitions \cite{Hawkins2015,Vampa2014}.

\begin{figure}[h]
\centering
\includegraphics[width=0.48\textwidth]{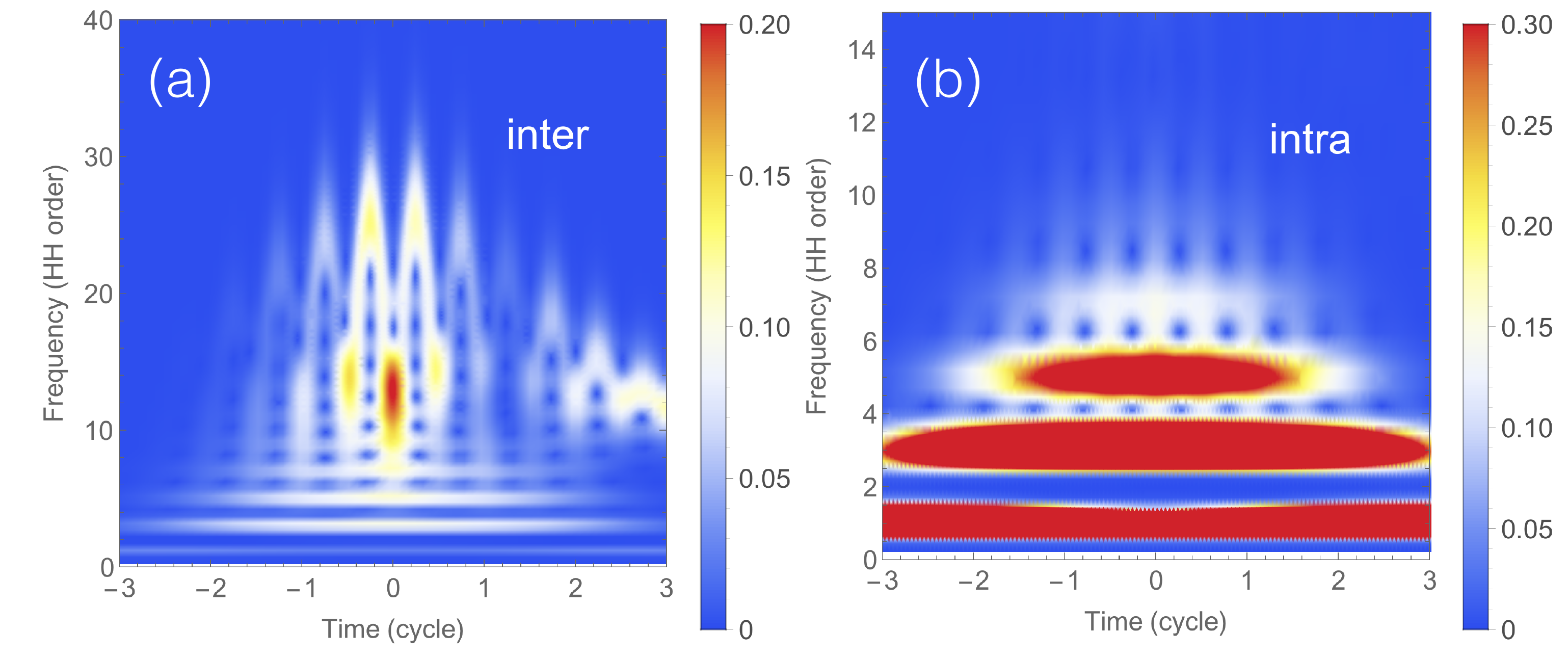}
\caption{(Color online) The wavelet transform (linear scale) of the (a) inter-band  and (b) intra-band current of a Houston model. The harmonics generated from inter-band dynamics mainly happens at the peak of the vector potential and there is a clear chirp, while the harmonics generated from intra-band dynamics mainly happens at the peak of the field and don't have clear chirp. The yield is saturated for the lowest frequencies in (b).}
\label{fig:wavelet_inter_intra}
\end{figure}

For the remainder of this section, we return  to the lower intensity case in which we observe a clear distinction between inter- and intra-band dynamics. In order to investigate the electron temporal dynamics using the Houston basis wave function, we perform a wavelet transform of the intra- and inter-band currents. The wavelet transform is similar to a windowed Fourier transform and provides  time-frequency information about f the two contributions. In the wavelet transform we use an order 10 Gabor wavelet to achieve a balanced resolution in both the time and frequency domains. Fig.~\ref{fig:wavelet_inter_intra} shows the resulting time-frequency profiles for the two different contributions, which clearly exhibit two distinct types of dynamics. We will discuss these separately in the following.

The time structure of the inter-band current in Fig.~\ref{fig:wavelet_inter_intra}(a) exhibits two emission times that are symmetrically placed around the peak of the vector potential in each half cycle. These two emission time profiles have opposite chirps, and are merged at the cutoff frequency. In the momentum picture we described above, these two emission times arise from the fact that the lattice momentum $k(t)$, and thereby the time-dependent band gap $\varepsilon(k(t))$, traverses all allowed values twice in a half cycle. Since the two emission times contribute to the current with about the same strength, our model suggests that one or the other must be filtered out to obtain a train of identical attosecond pulses from the plateau.  In contrast, harmonics near the cutoff frequency are generated near the time when the band gap $\varepsilon(k(t))$ is the largest, corresponding to the peaks of the vector potential in our case. 

The time structure of the intra-band current Fig.~\ref{fig:wavelet_inter_intra}(b), on the other hand, follows simply from the time-dependence of the band structure $\varepsilon(k(t))$, {\it i.e.}, the curvature of the bands.  The emission times correspond to those times when the  band has the largest curvature (the largest rate of change  in the group velocity), which corresponds to the zeroes of the vector potential, as shown in Fig.~\ref{fig:wavelet_inter_intra}(b). The time-frequency features of the intra-band current can also be seen from the cosine expansion of the conduction band in \cite{Ghimire2010}
\begin{equation}
j_{\text{intra}}(t) = \sum_{s=1}^\infty D_s \sin\left[(2s-1)\omega t\right]
\end{equation}
where $D_s$ is related to the band structure and the strength of the field. From this expansion, it is clear that all the harmonics are generated at the same time and there is no chirp in the generated field. We propose that the differences in the two time-frequency characteristics could be used as an experimental signature of the intra- and inter-band dynamics.


An analogy can be drawn between the Houston picture and a strongly driven two-level system to help understand the picture of the inter- and intra-band dynamics that naturally emerges from it. In the two-level system, the adiabatic states are the instantaneous eigenstates of the time-dependent Hamiltonian. The dynamics of the system can then be separated into an adiabatic part and a diabatic part using the adiabatic states as the basis. The adiabatic motion describes the evolution of the system  along the adiabatic states, whereas the diabatic motion describes the transition between the adiabatic states. 
In solids, this same separation is achieved in the Houston states, which are the instantaneous eigenstates (adiabatic states) of the system, as shown in Eq.~\eqref{eqn:Houston_state_as_adabatic_state}. Since the adiabatic states are time-dependent themselves, the adiabatic evolution  generates nontrivial dynamics by itself  (the Bloch oscillation, see Appendix B). Similarly, the inter-band dynamics can also be understood as  diabatic transitions between  adiabatic states, the same as in a two-level system.
Note that besides these two descriptions, a third commonly used description for a two-level system is  the Floquet states, which are the true eigenstates of the laser-dressed system. The counterpart for this description in a solid is the Floquet-Bloch theory and it has been well studied in \cite{Faisal1997, Hsu2006}.

\section{Numerical difficulty of the Houston basis}

 As a matter of practice, numerical models are more trustworthy 
if convergence can be achieved with respect to all of the parameters in the model. In the present case, the number of bands would seem to be such a parameter, along with the number of $k$ points and time step size. In  calculations using the Houston basis, however, we find that the numerical results converge poorly if we include more than three bands in our model. In this section we discuss the numerical difficulty of including more bands in the Houston basis, which raises questions about the validity of the Bloch oscillation picture for higher bands.

\begin{figure}[h]
\centering
\includegraphics[width=0.48\textwidth]{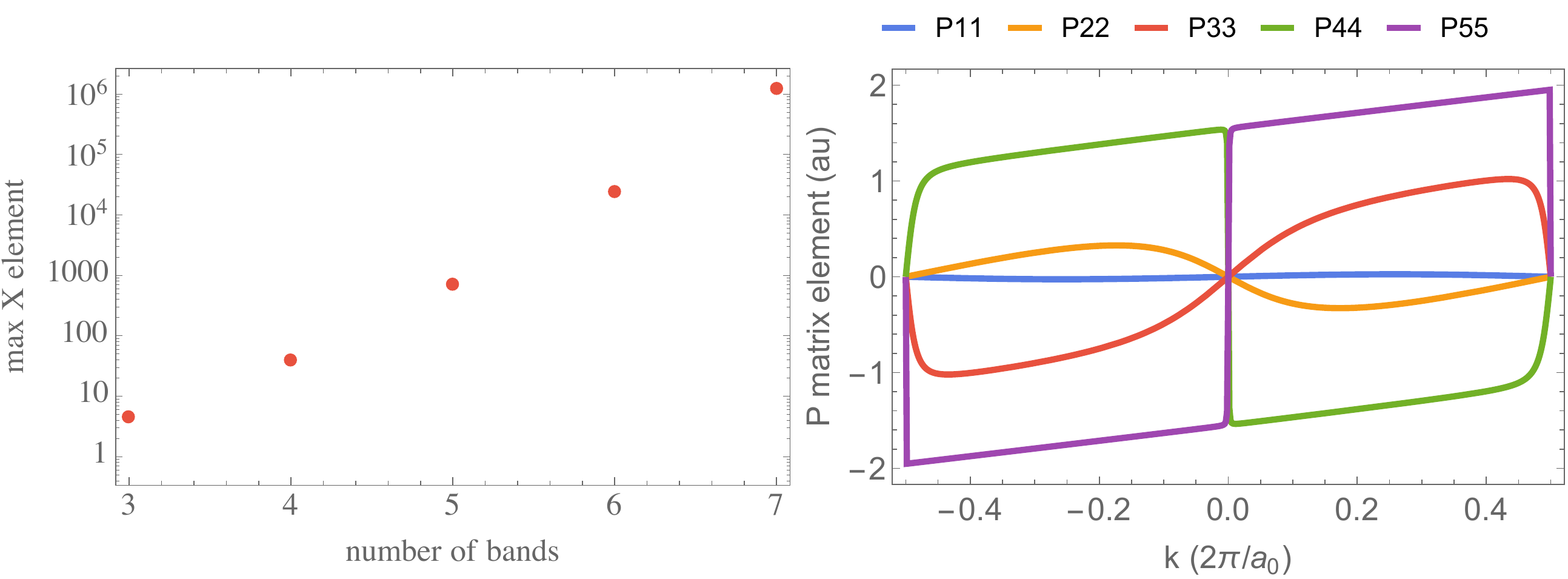}
\caption{(Color online)  The maximum matrix element of the X matrix seems to increase rapidly as we include more bands. The diagonal part of the P matrix is approaching discontinuous as the bands get higher. The parameters used here is the same as in Fig.~\ref{fig:Multiband_spectrum}. } 
\label{fig:X_mtx_exp_grow_P_mtx_discontinuous}
\end{figure}

As we include more bands in the Houston model, the gap between neighboring bands becomes very small, and the $X$ matrix elements in Eq. (\ref{eqn:X_matrix}) increase rapidly. In fact, the $X$ matrix grows approximately exponentially as a function of the number of bands for our current parameters, as shown in Fig.~\ref{fig:X_mtx_exp_grow_P_mtx_discontinuous}(a).
Since the X matrix is responsible for the inter-band transitions, in order to have the TDSE numerically converged, we must have a time step that is small enough to resolve its largest matrix element. Apart from resolving the $X$ matrix, the time step must also be small enough to resolve the largest  gap energy in the band structure. As we increase the number of bands, the requirement by the $X$ matrix of the time step becomes the dominant one. For example, to include 7 bands, the $X$ matrix requires a million time steps per laser cycle, which is much more than that required to resolve the highest energy difference in the system.

Another numerical difficulty in the Houston basis is the near discontinuity in the momentum operator matrix elements for higher bands. As we go to higher bands, the band structure will have sharp turning points at the band center and band edge. These turning points result in an almost discontinuous behavior of the momentum operator matrix element as shown in Fig.~\ref{fig:X_mtx_exp_grow_P_mtx_discontinuous}(b). This discontinuous behavior can also be expected from the relationship between the momentum operator and the derivative of the band structure Eq.~\eqref{eqn:relation_band_p_operator}. 

These numerical difficulties suggest that separation of the inter-band and intra-band dynamics may not be an optimal picture for an electron in the higher bands where Zener tunneling between the neighboring bands is so large that an electron never performs solely intra-band motion. Instead, it tunnels through the avoided crossing almost like a free electron \cite{Hawkins2015}. The artificial separation  of a Bloch oscillation motion from the almost free electron motion  complicates the physical picture as well as makes the numerical calculation difficult. This separation is the underlying cause of the problematic behaviors of the $p$ and $X$ matrix shown in Fig.~\ref{fig:X_mtx_exp_grow_P_mtx_discontinuous}. This is also a general problem for studying strongly driven systems using an adiabatic basis. In these systems, the avoided crossings between the adiabatic states are so small that the transitions between them blow up. 
In those conditions, the true eigenstates of the system (the Floquet states), may be a better basis to work with, though it is difficult to apply these methods to broadband driving pulses.

\section{summary}

We have studied  high harmonic generation in a model transparent solid using a 1D single-electron model and found that this system presents rich nonlinear dynamics. In the laser induced current, we see a high harmonic spectrum with multiple plateaus. Using the numerically robust Bloch  basis, we have shown that the primary plateau is due to transitions between the valence band and the lowest conduction band, whereas the secondary plateau and more generally higher frequencies in the spectrum are due to contributions from higher lying bands. We find that the cutoff of the primary plateau scales linear with the field strength, in agreement with current experiment \cite{Ghimire2010}, and we predict that this cutoff also scales linearly with the driving wavelength. 

We have also shown that the dynamics of our model system can be expressed in either the Bloch  basis or the Houston  basis, and solutions in these two basis are connected through a unitary transform. The Houston  basis allow for an intuitive separation of intra- and inter-band dynamics, and we found that for moderate intensities the harmonic radiation is due primarily to inter-band dynamics, in agreement with the prediction in \cite{Vampa2014}. At higher intensities, though, this artificial separation becomes more problematic as more bands are strongly coupled to each other which manifests itself in the numerical calculation becoming unstable. By limiting the dynamics in the Houston basis to just the conduction band, we found that for these high intensities, an alternative interpretation of the extended spectral range is provided by driven conduction-band Bloch oscillations that traverse the entire Brillouin zone. Our Houston basis calculations also suggested that, in general, the {\it cutoff} in the harmonic spectrum is tied to inter-band dynamics through the time-dependence of the population in the valence and conduction bands. 

Finally, we showed that in the regime where the intra- and inter-band dynamics can be clearly separated, they have very different time-frequency signatures, and we proposed that this could be harnessed to experimentally characterize the harmonic generation dynamics.

\begin{acknowledgments}
At LSU, this work was supported by the National Science Foundation under Grant No. PHY-1403236. At Stanford/SLAC, this work was supported by the U.S. Department of Energy, Office of Science, Office of Basic Energy Sciences, through the AMOS program within the Chemical Sciences Division (DAR) and the Office of Science Early Career Research Program (SG). High performance computational resources were provided by the Louisiana State University High Performance Computing Center.
\end{acknowledgments}

\appendix

\section{The unitary transformation of solutions in the Bloch and Houston basis}

In this section, we show that the solution of the TDSE in the Bloch state basis is connected to that in the Houston basis by a unitary transformation. In the Bloch state approach, we solve the TDSE in the velocity gauge, and the equation reads
\begin{equation}
i\hbar \frac{\partial}{\partial t}\ket{\psi^B_{k_0}(t)}=\left[\frac{p^2}{2m}+V(x)+\frac{eA(t)}{m}p\right]\ket{\psi^B_{k_0}(t)},
\end{equation}
where $k_0$ labels the single lattice momentum channel that we consider, since different channels are independent. We then express the solution in the basis of Bloch states, which are the eigenstates of the field-free Hamiltonian
\begin{equation}
\ket{\psi^B_{k_0}(t)}=\sum_n c_n(t) \ket{\phi_{nk_0}},
\label{eqn:wavefunction_in_Bloch_basis}
\end{equation}
where $c_n$'s are the time-dependent energy band amplitudes that this model solves for.

In the Houston approach, TDSE reads
\begin{equation}
i\hbar \frac{\partial}{\partial t}\ket{\psi^H_{k_0}(t)}=\left[\frac{\left(p+eA(t)\right)^2}{2m}+V(x)\right]\ket{\psi^H_{k_0}(t)}.
\label{eqn:Houston_TDSE_appendix}
\end{equation}
The wave function is expressed in the Houston states
\begin{equation}
\ket{\psi^H_{k_0}(t)}=\sum_n a_{nk_0}(t) \ket{\widetilde\phi_{nk_0}(t)},
\label{eqn:wavefunction_in_Houston_basis}
\end{equation}
The Houston states are related to the Bloch states by 
%
%
\begin{align}
\ket{\widetilde\phi_{nk_0}(t)}&=e^{-ie\hat xA(t)/\hbar}\ket{\phi_{nk(t)}}.
\label{eqn:relation_Houston_to_Bloch}
\end{align}
%
%
%
We can then match the wave function in these two basis, and get the unitary transformation matrix between the expansion coefficients. Since $\ket{\psi^B_{k_0}}$ and $\ket{\psi^H_{k_0}}$ satisfy the Schr\"odinger's equations that differ by a time-dependent  $A^2$ term, they must be related by a phase factor
\begin{equation}
\ket{\psi^B_{k_0}(t)}=e^{-ie\Lambda/\hbar}\ket{\psi^H_{k_0}(t)}
\end{equation}
where
\begin{equation}
\Lambda=-\frac{e}{2m}\int_0^t A^2(t')dt'.
\end{equation}
%
%
%
Substitude Eq.~\eqref{eqn:wavefunction_in_Bloch_basis} and Eq.~\eqref{eqn:wavefunction_in_Houston_basis}, we finally come to
\begin{equation}
a_{nk_0}(t)=e^{ie\Lambda/\hbar}\sum_{n'} c_{n'k_0}(t) \braket{\phi_{nk(t)}|e^{ie\hat xA(t)/\hbar}|\phi_{n'k_0}}
\end{equation}
which is the unitary transformation we are seeking. Note that this unitary transformation is very similar to the Kramers-Henneberger transformation in the atomic case, where a spatial transformation shifts the system into the accelerated frame, corresponding to the motion of a charged particle in the electric field \cite{Grossmann2008}. The Kramers-Henneberger transformation is a transformation in space whereas the transformation here is in momentum space.

\section{Connection of the single band model to the intra-band motion}

In this section we show that if we prevent the inter-band dynamics, we essentially come back to the single conduction band model used in \cite{Ghimire2010,Ghimire2012}.

If the inter-band dynamics is not allowed, then the inter-band transition matrix $X$ vanishes in Eq.~\eqref{eqn:EOM_Houston}, and the population on each band stays the same as the initial condition. The solutions for Eq.~\eqref{eqn:Houston_TDSE_appendix} are exactly the Houston states apart from a phase
\begin{align}
a_{nk_0} (t) &= a_{nk_0} (0) e^{-\frac{i}{\hbar}\int_0^t \varepsilon_n(k(t')) dt'}\\
\ket{\psi^H_{k_0}(t)} &= e^{-i eA(t) \hat x/\hbar} \sum_n a_{nk_0} (t) \ket{\phi_{nk(t)}}.
\end{align}
Substituting $a_{nk_0}$ into Eq.~\eqref{eqn:Houston_total_current} the total current is
\begin{equation}
j_{\text{tot}} =  -\frac{e}{m}  \sum_{n} \left| a_{nk_0} (0)\right|^2 \braket{\phi_{nk(t)}|\hat p|\phi_{nk(t)}}.
\end{equation}
If we consider the same situation as in \cite{Ghimire2010,Ghimire2012} where initially only the lowest conduction band is populated and other bands are empty, then the total current reduces to
\begin{equation}
j_{\text{tot}} =  -\frac{e}{m}  \braket{\phi_{ck(t)}|\hat p|\phi_{ck(t)}} .
\label{eqn:Houston_current_relate_to_p_operator}
\end{equation}
 This then reduces to the single conduction band model used in \cite{Ghimire2010,Ghimire2012}, where the semi-classical current is derived from a group velocity
\begin{equation}
j_{\text{tot}}=-ev_g=-\frac{e}{\hbar} \left. \frac{\pd \varepsilon_c}{\pd k}\right|_{k=k(t)},
\end{equation}
since the diagonal elements of the momentum operator is related to the band structure by
\begin{equation}
\braket{\phi_{nk}|\hat p|\phi_{nk}} = \frac{m}{\hbar}\frac{\partial \varepsilon_n}{\partial k}.
\label{eqn:relation_band_p_operator}
\end{equation}
The proof of the last step can be found in many textbooks. For example, see chapter III in \cite{Mott1958} or Appendix E in \cite{Ashcroft1976}.

\bibliography{library.bib}

\end{document}